# Augmented Sequence Impedance Networks of Grid-tied Voltage Source Converter for Stability Analysis[1]

Chen Zhang[1], Xu Cai[1], Atle Rygg[2] and Marta Molinas[2]

1 Department of Wind Power Research Center , Shanghai Jiao Tong University, State Energy Smart Grid R&D Center B201,Dong Chuan Road No.800, 200240,Shanghai.

2 Department of Engineering Cybernetics, O.S. Bragstads plass 2, Elektro D*244, NO-7491,Trondheim

Corresponding author: Chen Zhang

*nealbc@sjtu.edu.cn; or zhangchencumt@163.com

*Abstract*—Impedance-based stability analysis normally splits an interconnected system into source and load subsystems and the stability of the closed loop system can be analyzed by the combined input-output gains of source and load. This is an intuitive approach in the case of Single-Input Single-Output (SISO) systems. However, in the case of grid-tied voltage source converter (VSC) systems, *dq* impedances of source and load (VSC) subsystems are typically Multi-Input Multi-Output (MIMO) systems in which case the Generalized Nyquist Criterion (GNC) is required for analyzing the closed loop stability, which increases the complexity of the analysis. Modeling the impedances in phase domain has been an alternative to remove the coupling due to passive elements e.g. inductance and capacitance. However, recent research has shown that also in the sequence domain there still exist couplings between sequences and they can be important for stability analysis, especially in the low frequency range. Therefore, it seems evident that matrix impedance models are required whenever accuracy is the first priority.

This paper explores further the coupling between positive and negative sequence impedances, in particular the dependency and bindings between them. It shows that the couplings in each subsystem can be compounded into two non-coupled sequence impedances if the source and load subsystems are viewed as an integrated system instead of as two separate subsystems. Therefore, in this paper the closed loop gain is used instead of the minor loop gain for stability analysis. Subsequently, the closed loop gain is analytically derived using *dq* sequence impedances and two decoupled SISO systems are obtained which are defined as Augmented Sequence Impedance Networks (ASIN). From this, the stability analysis of the closed loop system can be performed directly on the loop impedance of ASIN with the principal of Argument. In order to justify the validity and accuracy, both numerical and time domain simulations are presented. In addition, the impact of PLL on sequence couplings and circuit dynamics are elucidated.

**Nomenclatures and Notions:**

Phase Locked Loop— PLL

Augmented Sequence Impedance—ASI

Augmented Sequence Impedance Networks—ASIN

Generalized Nyquist Criterion —GNC

---

[1] This paper has been submitted to IEEE JESPTPE on (21-Nov-2016) and has been denied publication due to lack of experimental proof on (09-Mar-2017)

Argument Principle —AP

Intrinsic Oscillatory Point —IOP

*u* denotes the voltage

*i* denotes the current

*s* denotes Laplace operator

ω denotes angular speed

*δ* is the angle displacement

*Z* and *Y* denote impedance and admittance

Variables in bold denotes vector or complex-valued

*pll* in superscript denotes variables in PLL domain

*d* and *q* in subscript denotes *d* and *q* components

*p* and *n* in superscript denotes positive and negative sequence variables

*c* in subscript denotes variables of converters

*s* in subscript denotes variables of grid

*l* in subscript denotes loop

∗ in superscript denotes conjugate operator.

## I. INTRODUCTIONS

Modeling a complex interconnected system into source and load impedance is an effective approach for small signal stability analysis. Hence stability of the closed loop system can be analyzed by the combined source and load gains [1]. This is an intuitive approach in case of Single-Input and Single-Output (SISO) systems. In the case of grid-tied voltage source converter (VSC) systems, impedances can be modeled either in synchronously rotating frame (*dq* domain) or in stationary frame (phase domain). In *dq* domain, VSC becomes a nonlinear time-invariant system and can be linearized in a simple way. It was first performed in [2], later by [3] and [4]. In phase domain, harmonic linearization in [5] is adopted due to its time-variant nature. Since impedances given by both methods are typically in matrix forms, the minor loop gains are Multi-Input and Multi-Output (MIMO) systems and the Generalized Nyquist Criterion (GNC) [6] is required for stability analysis. In order to use the classic Nyquist Criterion in [7], two decoupled sequence impedances can be obtained in [5] if the couplings between sequences are neglected partially.

On the other hand, recent research has shown that *dq* impedance can be transformed into sequence impedance by matrix manipulation in [8] and [9]. And in [9], the condition for when the sequence domain impedance matrix is decoupled is defined as Mirror Frequency Decoupled (MFD) system. According to this condition, VSC with Phase-locked-loop (PLL) and current vector controller does not constitute a MFD system, leading to couplings between sequences which impacts on the stability as stressed in [10] and [11].

This paper further explores the couplings between positive and negative sequence impedances, in particular the dependency and bindings between them. The primary objective of the paper is to develop two decoupled SISO models for stability analysis, which have the same accuracy as the typical MIMO matrix models. In Section II, a thorough analysis of the origin and the propagation of sequence couplings are given first. It is then explained why the paper focuses on closed loop gain instead of minor-loop gain. In Section III, sequence impedance modeling in *dq* domain is presented in detail. Subsequently, the closed loop gain is calculated using the derived source and load *dq* sequence impedances in Section IV. Then, two decoupled SISO systems defined as Augmented Sequence

Impedance Networks (ASIN) in this paper are finally obtained, and stability analysis is performed by using the loop impedance of ASIN with the use of Argument Principle. Next in Section V and VI, in order to justify the validity and accuracy of ASIN for stability analysis, both numerical and time domain simulations are presented. In addition, the impact of PLL on sequence couplings and circuit dynamics are elucidated in detail.

## II. INTRODUCTION TO SEQUENCE COUPLINGS

There are three domains being considered in the analysis of this paper, phase domain in stationary frame, *dq* domain and phase domain in *dq* frame. Sequence impedances in *dq* frame can be obtained by applying symmetrical decomposition method on *dq* impedances as shown in Fig.1(c). Since the load (converters) does not constitute a MFD system [9], sequence impedances in *dq* frame are coupled, leading to coupled positive and negative sequence responses regardless of the types of inputs. This unbalanced behavior of sequence impedances in *dq* frame will further give rise to the coupled sequence impedances in stationary frame.

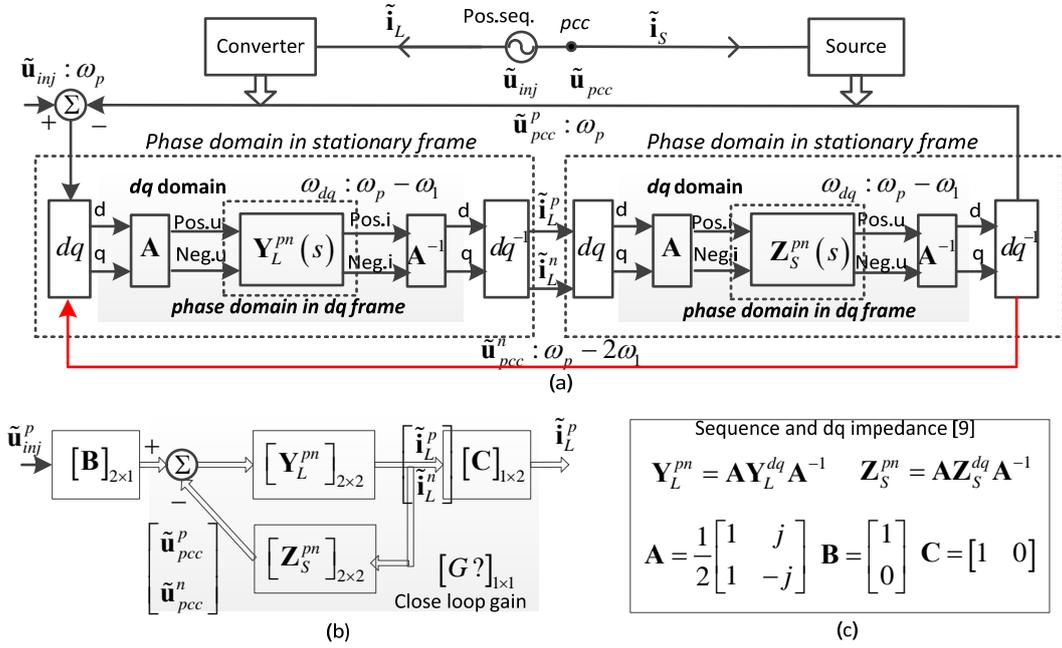

Fig.1. Small signal propagation between different domains

According to the above analysis, a systematic view of the occurrence of sequence couplings in phase domain (stationary frame) is shown in Fig.1 (a). The small signal positive sequence voltage perturbation propagates from the load to the source and then feedbacks to the inputs forming a closed loop system. A more compact description of this procedure by control blocks is shown in Fig.1 (b). Obviously, it's a typical MIMO feedback system and its stability can be studied by the minor loop gain using the GNC. In this case, there is no difference between *dq* impedance and sequence impedance since both of them are matrices and the GNC are required. Hence, it somehow contradicts the main benefit of applying sequence impedance.

This paper will move a step further in this direction by finding the closed loop gain in Fig.1 (b) analytically. As a consequence, a decoupled positive and negative sequence network will be identified.

## III. SEQUENCE IMPEDANCE OF GRID-TIED VSC IN PLL DOMAIN

Dq impedance can be modeled by transforming variables from converter *dq* domain into system *dq* domain [12]. This can be complicated due to VSC controllers usually have larger numbers of variables than the grid variables. In this work, the converter *dq* domain which is redefined as PLL domain is adopted for modelling and analyzing, that is, variables are transformed from system *dq* domain into PLL domain.

B. *dq impedance modeling of grid-tied VSC in PLL domain*

Fig.2 gives the grid-tied VSC system analyzed in the paper. The system contains a VSC, medium voltage step up transformer (usually used in wind power integration) and a Thevenin equivalent grid.

Circuit voltage equation in PLL domain can be written as below:

$$\mathbf{u}_c^{pll} = L_\Sigma \frac{d\mathbf{i}_c^{pll}}{dt} + \left(R_\Sigma + j\omega_{pll}L_\Sigma\right)\mathbf{i}_c^{pll} + \mathbf{u}_s^{pll} \quad (1)$$

$R_\Sigma = R_{filter} + R_T + R_s$, $L_\Sigma = L_{filter} + L_T + L_s$ are the total resistance and inductance in circuit (filter, transformer and equivalent grid impedance). The capacitance in the output LCL-filter is disregarded since its harmonic oscillation frequency is significantly higher than frequencies of interest in this paper [13].

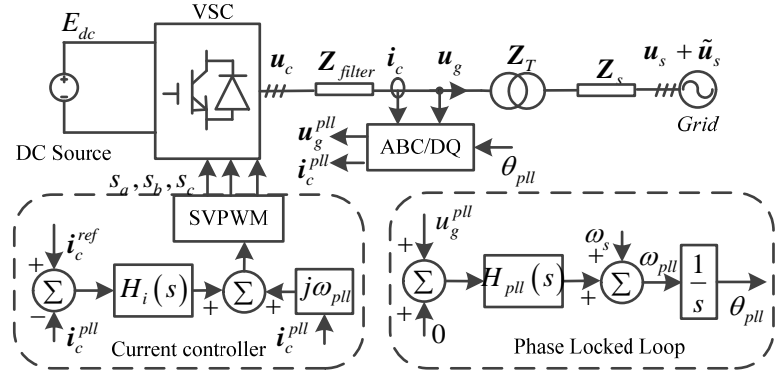

Fig.2 Schematic of grid-tied VSC system

Linearizing (1) around the stationary point $\left(\mathbf{I}_{c0}^{pll}, \mathbf{U}_{s0}^{pll}\right)$, and taking Laplace transform gives:

$$\tilde{\mathbf{u}}_c^{pll}(s) = \mathbf{Z}_\Sigma(s)\tilde{\mathbf{i}}_c^{pll}(s) + j\tilde{\omega}_{pll}(s)L_\Sigma \mathbf{I}_{c0}^{pll} - j\mathbf{U}_{s0}^{pll}\tilde{\delta}_{pll}(s) + \tilde{\mathbf{u}}_s^{pll} \quad (2)$$

$\tilde{\mathbf{u}}_c^{pll}(s) = \tilde{u}_{cd}^{pll}(s) + j \cdot \tilde{u}_{cq}^{pll}(s)$ is complex-valued variable in Laplace domain. For simplicity, the Laplace operator is omitted hereafter. $\mathbf{Z}_\Sigma(s) = L_\Sigma(s + j\omega_s) + R_\Sigma$ is the total impedance seen from the converter terminal ($\mathbf{u}_c$). $\tilde{\delta}_{pll}(s)$ is the angle displacement between grid and PLL output.

Accordingly, the linearized current controller and PLL model can be derived:

$$\tilde{\mathbf{u}}_c^{pll} = \mathbf{H}_i(s) \cdot \left(\tilde{\mathbf{i}}_c^{ref} - \tilde{\mathbf{i}}_c^{pll}\right) \quad (3)$$

$$\tilde{\delta}_{pll} = \mathbf{G}_{pll}(s) \cdot \tilde{u}_{cq}^{pll} \quad (4)$$

In (4) the node voltage for PLL sampling point ($\tilde{\mathbf{u}}_g^{pll}$) is an intermediate variable and can be eliminated by applying Kirchhoff laws on the circuit. ($\mathbf{H}_i(s), \mathbf{G}_{pll}(s)$ are given in Appendix. A.).

$\tilde{\mathbf{i}}_c^{ref}$ is considered as zero since only current controller applied.

B. *Transforming dq impedance into dq Sequence impedance*

*dq* impedance can be viewed as a two phase system in frequency domain, and symmetrical decomposition method can be applied to find the sequence impedance in the same domain.

For example, $\tilde{\mathbf{u}}_c^{pll}$ can be decomposed at frequency $s = j\omega$ as follows:

$$\tilde{\mathbf{u}}_c^{pll}(j\omega) = \frac{\tilde{\mathbf{u}}_{cd}^{pll} + j\tilde{\mathbf{u}}_{cq}^{pll}}{2}(j\omega) + \left(\frac{\tilde{\mathbf{u}}_{cd}^{pll} - j\tilde{\mathbf{u}}_{cq}^{pll}}{2}\right)^*(-j\omega) \quad (5)$$

In (5), the positive and negative sequence component can be written in matrix form as in [9]:

$$\begin{bmatrix} \tilde{\mathbf{u}}_c^p \\ \tilde{\mathbf{u}}_c^n \end{bmatrix} = \frac{1}{2}\begin{bmatrix} 1 & j \\ 1 & -j \end{bmatrix}\begin{bmatrix} \tilde{\mathbf{u}}_{cd}^{pll} \\ \tilde{\mathbf{u}}_{cq}^{pll} \end{bmatrix} \quad (6)$$

Same procedure can be used to decompose current and PLL, and gives:

$$\tilde{\mathbf{i}}_c^{pll} = \tilde{\mathbf{i}}_c^p + \left(\tilde{\mathbf{i}}_c^n\right)^* \quad (7)$$

$$\tilde{\boldsymbol{\delta}}_{pll} = \mathbf{G}_{pll}(s) \cdot \frac{\left(-j\tilde{\mathbf{u}}_c^p + j\tilde{\mathbf{u}}_c^n\right) + \left(-j\tilde{\mathbf{u}}_c^p + j\tilde{\mathbf{u}}_c^n\right)^*}{2} \quad (8)$$

Insert (5), (7), (8) into (2) and rewritten it in matrix form, gives the coupled sequence impedance:

$$\begin{bmatrix} \tilde{\mathbf{u}}_c^p \\ \tilde{\mathbf{u}}_c^n \end{bmatrix} = \begin{bmatrix} \mathbf{Z}_{grid}^p & \\ & \mathbf{Z}_{grid}^n \end{bmatrix}\begin{bmatrix} \tilde{\mathbf{i}}_c^p \\ \tilde{\mathbf{i}}_c^n \end{bmatrix} + \begin{bmatrix} 0 & \mathbf{D}_{pll} \\ \mathbf{D}_{pll}^* & 0 \end{bmatrix}\begin{bmatrix} \tilde{\mathbf{u}}_c^p \\ \tilde{\mathbf{u}}_c^n \end{bmatrix} + \begin{bmatrix} \tilde{\mathbf{u}}_s^p \\ \tilde{\mathbf{u}}_s^n \end{bmatrix} \quad (9)$$

$$\begin{bmatrix} \tilde{\mathbf{u}}_c^p \\ \tilde{\mathbf{u}}_c^n \end{bmatrix} = -\begin{bmatrix} \mathbf{Z}_c^p & \\ & \mathbf{Z}_c^n \end{bmatrix}\begin{bmatrix} \tilde{\mathbf{i}}_c^p \\ \tilde{\mathbf{i}}_c^n \end{bmatrix} \quad (10)$$

$\mathbf{Z}_{grid}^p$ $\mathbf{Z}_{grid}^n$ $\mathbf{Z}_c^p$ $\mathbf{Z}_c^n$ $\mathbf{D}_{pll}$ are complex-valued transfer functions given in Appendix A. $\tilde{\mathbf{u}}_s^p, \tilde{\mathbf{u}}_s^n$ is the injected perturbation of sequence voltages. Worth to mention is that the conjugate operator only acts on complex values and not on the Laplace operator (E.g., $\mathbf{Z}_\Sigma^* = R_\Sigma + (s - j\omega_s)L_\Sigma$).

C. *Sequence networks and its properties*

The positive and negative sequence networks can be drawn in Fig.3 based on (9) and (10). Since the modeling domain is moved to the PLL, the source seen from the PLL frame is unbalanced, and the sequence networks are coupled with two mutually dependent voltage sources. In addition, the coupling strength can be measured by the complex-valued gain $\mathbf{D}_{pll}$ (Appendix A), which is a function of PLL bandwidth.

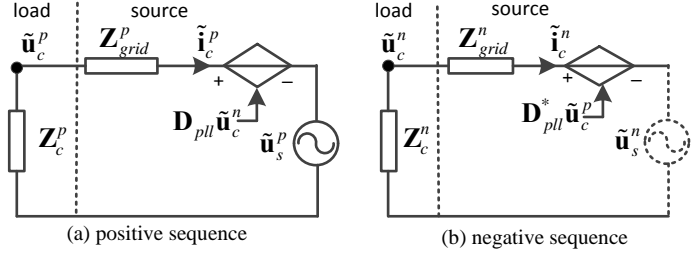

(a) positive sequence  (b) negative sequence

Fig. 3 Sequence networks

If the system is perturbed at the frequency far beyond the bandwidth of PLL, then $\mathbf{D}_{pll}$ will approach zero, and positive and negative sequence network will be decoupled. In other words, decoupled sequence impedance can be used for stability analysis only if PLL bandwidth is sufficiently low. In addition, the dotted line in fig.3 (b) means when calculating the positive sequence network the negative sequence perturbation should be excluded, and vice versa.

IV. AUGMENTED SEQUENCE IMPEDANCE AND ITS NETWORKS

The positive and negative sequence network given in Fig.3 (a) has one dependent and one independent voltage source. Obviously, the sequence networks will be decoupled if the characteristics of the dependent voltage can be emulated by impedances. Hence, the Augmented Sequence Impedance (ASI) for positive and negative sequence is defined as:

$$\begin{cases} \mathbf{Z}_{couple}^{np} = \dfrac{\mathbf{D}_{pll}\tilde{\mathbf{u}}_c^n}{\tilde{\mathbf{i}}_c^p} \\ \mathbf{Z}_{couple}^{pn} == \dfrac{\mathbf{D}_{pll}^*\tilde{\mathbf{u}}_c^p}{\tilde{\mathbf{i}}_c^n} \end{cases} \quad (11)$$

Solve the negative sequence network in Fig.4 (b) for $\tilde{\mathbf{u}}_c^n$ and reinsert it into (11) gives:

$$\mathbf{Z}_{couple}^{pn} = -\frac{\left|\mathbf{D}_{pll}\right|^2 \left|\mathbf{Z}_c^p\right|^2}{\left|\mathbf{Z}_{grid}^p + \mathbf{Z}_c^p\right|^2} \cdot \left(\mathbf{Z}_c^p + \mathbf{Z}_{grid}^p\right) = -\mathrm{r}\left(\mathbf{Z}_c^p + \mathbf{Z}_{grid}^p\right) \quad (12)$$

And if the system is perturbed by negative sequence voltage, negative ASI is:

$$\mathbf{Z}_{couple}^{np} = -\frac{\left|\mathbf{D}_{pll}\right|^2 \left|\mathbf{Z}_c^p\right|^2}{\left|\mathbf{Z}_{grid}^p + \mathbf{Z}_c^p\right|^2} \cdot \left(\mathbf{Z}_c^n + \mathbf{Z}_{grid}^n\right) = -\mathrm{r}\left(\mathbf{Z}_c^n + \mathbf{Z}_{grid}^n\right) \quad (13)$$

With (12) and (13), the Augmented Sequence Impedance Networks (ASIN) can be drawn as Fig.4:

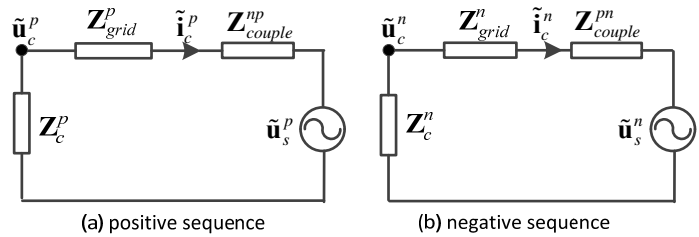

(a) positive sequence  (b) negative sequence

Fig. 4 Augmented sequence impedance networks

It is clearly shown in Fig.4 that sequence networks are decoupled and the effects of sequence couplings are included in ASI. Moreover, the impedance of ASIN is essentially the closed loop gain of the SISO system proposed in Fig.1 (b), and stability analysis of this system can be more convenient that the MIMO system.

The equations below have been generalized to cover also a system with sequence coupling in both subsystems. This is based on the block diagram in Figure 1. The derivation and result is presented in Appendix C.

## V. NUMERICAL STABILITY ANALYSIS

A. *Stability analysis of ASIN based on Argument principle*

Since the source and load system essentially form a MIMO feedback system as shown in Fig.1 (b), stability can be analyzed by its minor loop gain or the closed loop gain. However, in terms of closed loop gain, it is decoupled and can be represented by two SISO systems (positive and negative ASIN) with no accuracy lost. Therefore, stability of the closed loop system is equivalent to the loop impedance of ASIN.

Taking the positive ASIN as an example, the loop impedance is:

$$\mathbf{Z}_l^p = \mathbf{Z}_c^p + \mathbf{Z}_{grid}^p + \mathbf{Z}_{couple}^{pn} = (1-\mathbf{r})\left(\mathbf{Z}_c^p + \mathbf{Z}_{grid}^p\right) \quad (14)$$

And the negative ASIN loop impedance is $\mathbf{Z}_l^n = \left(\mathbf{Z}_l^p\right)^*$.

Obviously, if there are no right hand side poles of $\mathbf{Z}_l^p$ and $\mathbf{Z}_l^n$ in the complex plane, the system is stable. This can be formally justified by the Argument Principle (AP)[14] by which the locus of $\mathbf{Z}_l^p$ and $\mathbf{Z}_l^n$ do not clockwise encircle the (0,0j) point, if frequency vary from zero to positive infinite.

The next figures will compare AP (ASIN model) with GNC (*dq* phase domain matrix impedance, given in Appendix B) in terms of stability results. Parameters are given in Appendix D.

Conditions: Both locus of AP and GNC are plotted with different PLL bandwidth (from top to bottom down, the PLL bandwidth is 5Hz, 13Hz and 50Hz respectively). The grid impedance is calculated from the short circuit ratio (SCR), which is three in this case. Meanwhile, the current controller bandwidth is fixed to 200Hz, which is a typical value for grid-tied VSC with relatively low switching frequency (about 2.4 kHz).

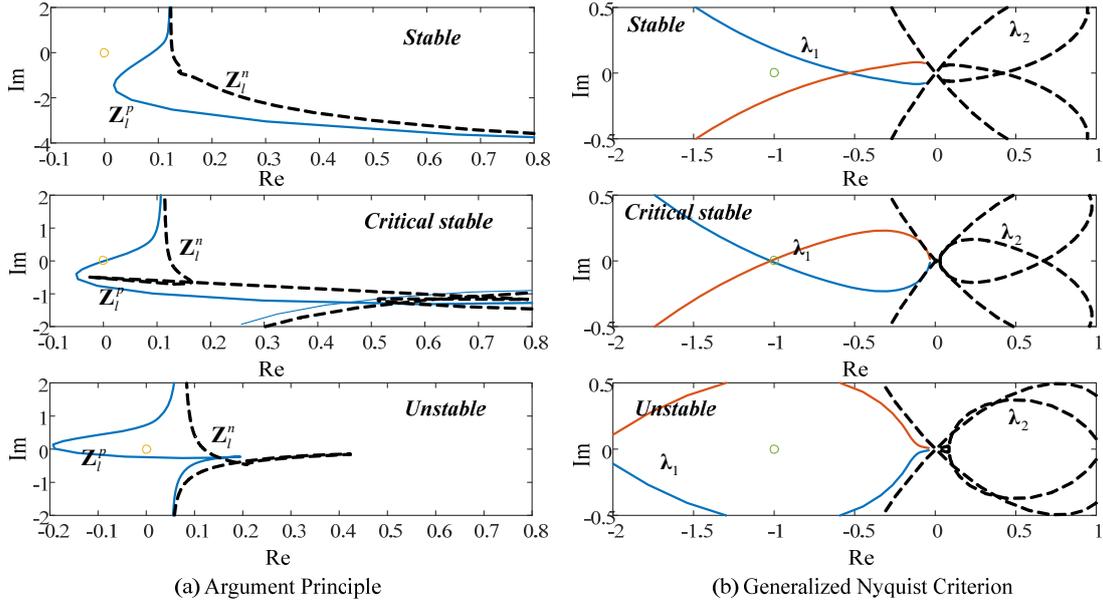

(a) Argument Principle  (b) Generalized Nyquist Criterion

Fig.5 Comparisons between AP and GNC on stability analysis

In Fig.6, it is clear that stability predicted by AP and GNC are same, especially at the critical stability point. On the other hand, stability margin of the system is reduced with increasing PLL bandwidth, as shown in Fig.6 (a) (from top to bottom). It implies that, in addition to making sequence networks coupled, the PLL might have some *negative resistance effects* on circuit dynamics.

B. *Physical interpretation of Argument Principle*

It has been known that the current controller (integral part) of VSC has capacitive effects on circuit dynamics while the grid is mostly inductive. Hence, the grid and VSC forms a resonant circuit with an Intrinsic Oscillatory Point (IOP). IOP can be calculated analytically by finding the frequency where the imaginary part of $\mathbf{Z}_l$ is zero. And damping can be evaluated by the real part of $\mathbf{Z}_l$ at IOP.

If PLL dynamics are not considered, the IOP is always stable due to the positive damping contribution from current controller (proportional part) and possible resistances in circuits. However, IOP and damping characteristics will change drastically when the effects of PLL are considered. As shown in (14), the loop impedance of ASIN is shaped by PLL with complex vector $1-\mathbf{r}$ in frequency domain. As a consequence, the system will be unstable if the shaped impedance has negative resistance at the new IOP, as explained in Fig.6. This is essentially the underlying meaning of Argument Principle.

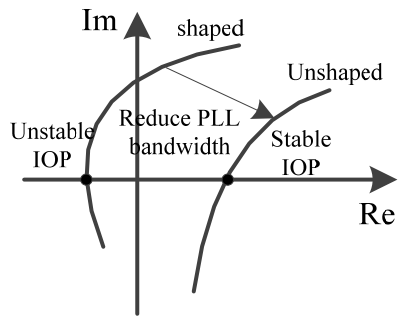

Fig 6 Physical interpretation of Argument Principle

Furthermore, PLL is basically a second order low pass filter, and any notable characteristics changes will only appear at frequency close to its bandwidth.. Hence, the IOP can be used as an approximation to the PLL bandwidth limit that will make the system unstable.

C. *Impacts of circuit resistance and SCR*

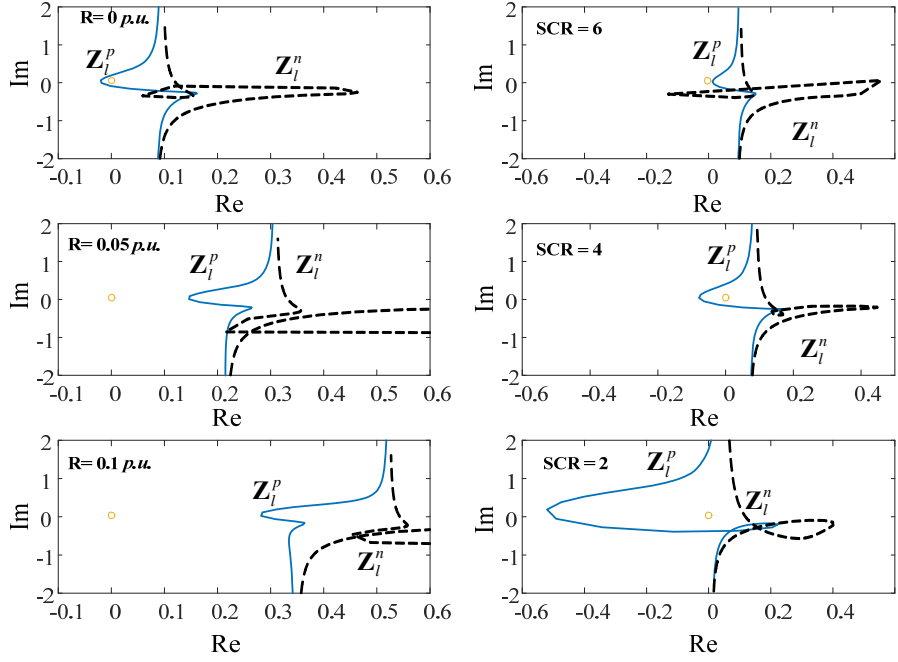

Fig. 7 Impacts of circuit resistance and SCR on stability

Fig.7 presents some AP plots with varying circuit resistance and SCR, where the bandwidth of PLL (50Hz) and current controller (200Hz) are kept constant. On the left column, system stability margin is enlarged with increasing circuit resistance. And on the right column, system stability margin is decreased with reducing SCR.

## VI. SIMULATIONS AND DISCUSSION

A. *Time domain simulations and verifications*

A test system shown in Fig.2 is built on PSCAD/EMTDC software with detailed switching model implemented. System parameters are given in Appendix D. Simulations are conducted by changing the PLL bandwidth with grid impedance (short circuit ratio is 3) and current controller bandwidth (200Hz) remaining constant. For the critical stable case in Fig.8, PLL bandwidth is 13Hz as shown in Fig.6, while for the unstable case, PLL bandwidth is chosen as 30Hz.

As shown in Fig.8 (a), both of the d and q axis currents oscillate with constant amplitude, indicating that the negative and positive resistance effects introduced by PLL and current controller are equal. Moreover, it proves that the bandwidth given by AP stability analysis fits well in time domain simulations. The second case shown in Fig.8 (b) also proves that if PLL bandwidth is further increased, the system will lose stability due to an undamped IOP. Same results are also predicted by numerical stability analysis in Fig.5.

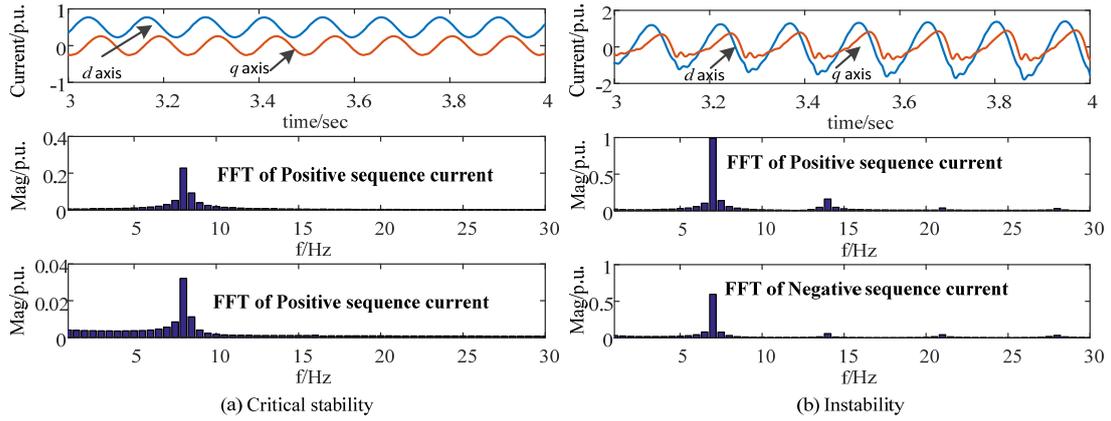

Fig. 8 Time domain simulation and analysis

B. *Discussions on the oscillatory behavior*

Although both cases in Fig.8 have oscillatory currents, the detailed behavior is not exactly the same. By performing FFT on currents waveforms, some important characteristics can be obtained. As shown in Fig.8 (FFT subfigures), both positive and negative sequence currents appeared, it proves that the PLL makes the sequence networks couple with each other. Moreover, the oscillation frequency of positive sequence current is the same as the negative one, but the magnitudes are different in both cases.

The ratio between negative and positive sequence current can be viewed as a measure of the unbalanced factor $f_u = \dfrac{\mathbf{I}_n}{\mathbf{I}_p}$. It is much smaller in Fig.8 (a) ($f_u \approx 0.15$) than in Fig.8 (b) ($f_u \approx 0.6$).

Therefore, the larger PLL bandwidth is the more unbalanced the system is, indicating more severe coupling effects of PLL on sequence networks.

## VII. CONCLUSIONS

Sequence couplings in individual source and load subsystem can be can be compounded into two non-coupled sequence impedances if the source and load subsystems are viewed as an integrated system instead of two separate subsystems.. Therefore, closed loop gain instead of minor loop gain is used for stability analysis. Consequently, two decoupled SISO systems defined as Augmented Sequence Impedance Networks (ASIN) are presented in this paper. Stability can be analyzed by plotting the loop impedances locus of positive and negative ASIN independently using the Principle of Arguments.

On the other hand, the PLL of current controlled VSC is the primary cause of the sequence couplings. It makes the positive and negative sequence network coupled by two PLL-dependent voltage sources. The lager the PLL bandwidth is, the more severe the coupling strength is. In addition, the PLL also has some negative resistance effects on circuit dynamics and may lead to undamped electrical oscillations. These conclusions can also be used as the guidance for VSC controller tuning especially for PLL bandwidth when VSC is synchronized to a weak grid.

## ACKNOWLEDGMENT

This paper and its related research are supported by grants from the Power Electronics Science and Education Development Program of Delta Environmental & Educational Foundation (DREM2016005).

APPENDIX:

A. *Key transfer functions used in the paper*

$$\mathbf{H}_i(s) = \left( k_p^c + \frac{k_i^c}{s} \right) \cdot L_{filter}$$

$$\mathbf{H}_{pll}(s) = \left(k_p^{pll} + \frac{k_i^{pll}}{s}\right)\frac{1}{U_s}$$

$$\mathbf{G}_{pll}(s) = \frac{\dfrac{\mathbf{H}_{pll}}{s}}{\dfrac{\mathbf{H}_{pll}}{s}k_m \cos\delta_{pll0}U_s + (k_m+1)}$$

$$\mathbf{C}_{pll}(s) = j\left(sL_\Sigma \mathbf{I}_{c0}^{dq} - \mathbf{U}_{s0}^{dq}\right)\mathbf{G}_{pll}(s)$$

$$k_m = \frac{L_{filter}}{L_s + L_T},\ \mathbf{Z}_{grid}^{p} = \frac{\mathbf{Z}_\Sigma(s)}{1 + j\dfrac{\mathbf{C}_{pll}(s)}{2}},\ \mathbf{Z}_{grid}^{n} = \left(\mathbf{Z}_{grid}^{p}\right)^*,$$

$$\mathbf{D}_{pll} = \frac{j\dfrac{\mathbf{C}_{pll}(s)}{2}}{1 + j\dfrac{\mathbf{C}_{pll}(s)}{2}},\ \mathbf{Z}_c^p = \mathbf{H}_i(s)\ \mathbf{Z}_c^n = \mathbf{H}_i^*(s).$$

B. *Stability analysis of ASIN based on Generalized Nyquist Criterion*

According to (9), the source (seen from converter terminal) coupled sequence impedance matrix is:

$$\mathbf{Z}_{source}^{pn} = \frac{1}{1 - \left|\mathbf{D}_{pll}\right|^2}\begin{bmatrix} \mathbf{Z}_{grid}^{p} & \mathbf{Z}_{grid}^{n}\mathbf{D}_{pll} \\ \mathbf{Z}_{grid}^{p}\mathbf{D}_{pll}^* & \mathbf{Z}_{grid}^{n} \end{bmatrix} = \gamma \cdot \begin{bmatrix} \mathbf{Z}_{grid}^{p} & \mathbf{Z}_{grid}^{n}\mathbf{D}_{pll} \\ \mathbf{Z}_{grid}^{p}\mathbf{D}_{pll}^* & \mathbf{Z}_{grid}^{n} \end{bmatrix} \quad \text{(A1)}$$

The converter sequence impedance matrix is:

$$\mathbf{Z}_{conv}^{pn} = \begin{bmatrix} \mathbf{H}_i(s) & \\ & \mathbf{H}_i^*(s) \end{bmatrix} \quad \text{(A2)}$$

And the minor loop gain of the interconnected system is $\mathbf{Z}_{source}^{pn} / \mathbf{Z}_{conv}^{pn}$. The complex eigenvalue of the loop gain can be calculated from the matrix determinant:

$$\det\left|\lambda\mathbf{I} - \frac{\mathbf{Z}_{source}^{pn}}{\mathbf{Z}_{conv}^{pn}}\right| \quad \text{(A3)}$$

And,

$$\begin{cases} \lambda_1 = \dfrac{1}{2}\left(\gamma\alpha + \gamma\alpha^* + \sqrt{\gamma^2\left(\alpha + \alpha^*\right)^2 - 4|\alpha|^2}\right) \\ \lambda_2 = \dfrac{1}{2}\left(\gamma\alpha + \gamma\alpha^* - \sqrt{\gamma^2\left(\alpha + \alpha^*\right)^2 - 4|\alpha|^2}\right) \end{cases} \quad \text{(A4)}$$

$\alpha = \dfrac{\mathbf{Z}_{grid}^{p}}{\mathbf{H}_i}$. According to GNC, stability of the system can be analyzed by the locus of these eigenvalues in the complex plane.

C. *The Generalized Augmented Sequence Impedance Networks (GASIN)*

Generalized ASIN is essentially the close loop gain of the system given in Fig.1 (b), that is:

$$\tilde{\mathbf{i}}_L^{pn} = \mathbf{C}\mathbf{Y}_L^{pn}\left(\mathbf{I} + \mathbf{Z}_s^{pn}\mathbf{Y}_L^{pn}\right)^{-1}\mathbf{B}\cdot\tilde{\mathbf{u}}_{inj}^{p} = \mathbf{G}_{1\times 1}\cdot\tilde{\mathbf{u}}_{inj}^{p} \quad (A5)$$

$$\mathbf{G} = \frac{1}{\mathbf{Z}_l^{p}\big|_G} \quad (A6)$$

Where $\mathbf{Z}_S^{pn} = \begin{bmatrix} \mathbf{Z}_s^{pp}(s) & \mathbf{Z}_s^{pn}(s) \\ \mathbf{Z}_s^{np}(s) & \mathbf{Z}_s^{nn}(s) \end{bmatrix}$, $\mathbf{Z}_L^{pn} = \begin{bmatrix} \mathbf{Z}_L^{pp}(s) & \mathbf{Z}_L^{pn}(s) \\ \mathbf{Z}_L^{np}(s) & \mathbf{Z}_L^{nn}(s) \end{bmatrix}$, $\mathbf{Z}_L^{pn} = \left(\mathbf{Y}_L^{pn}\right)^{-1}$. Matrix $\mathbf{C}$ and $\mathbf{B}$ are given in Fig1.(c).( L and S denotes load and source subsystems).

Insert (A5) into (A6):

$$\mathbf{Z}_l^{p}\big|_G = \mathbf{Z}_L^{pp} + \mathbf{Z}_S^{pp} - \frac{\left(\mathbf{Z}_L^{pn} + \mathbf{Z}_S^{pn}\right)\left(\mathbf{Z}_L^{np} + \mathbf{Z}_S^{np}\right)}{\mathbf{Z}_L^{nn} + \mathbf{Z}_S^{nn}} \quad (A7)$$

(A7) is the positive sequence GASIN. And the negative GASIN can be calculated in the same way:

$$\mathbf{Z}_l^{n}\big|_G = \mathbf{Z}_L^{nn} + \mathbf{Z}_S^{nn} - \frac{\left(\mathbf{Z}_L^{pn} + \mathbf{Z}_S^{pn}\right)\left(\mathbf{Z}_L^{np} + \mathbf{Z}_S^{np}\right)}{\mathbf{Z}_L^{pp} + \mathbf{Z}_S^{pp}} \quad (A8)$$

D. *Circuit parameters used in stability analysis and simulations*

| Name | Values | Name | Values |
|---|---|---|---|
| Nominal power (base power) | 2 MW | VSC filter inductance | 0.1 p.u. |
| Nominal voltage (base voltage) | 0.69kV | Transformer inductance(0.69/35kV) | 0.1 p.u. |
| DC link voltage | 1.1kV | Current reference | (0.5 +0j) p.u. |
| Fundamental frequency (base frequency) | 50Hz | Grid equivalent inductance | 0.33 p.u. (inverse of SCR) |

TABLE.I: Circuit parameters of the grid-tied VSC system